\documentclass[aps,pra,twocolumn,floatfix]{revtex4}
\usepackage{epsfig}
\usepackage{graphicx}
\usepackage{dcolumn}
\usepackage{amsthm,amsmath}

\begin{document}

\title{Comparing magic wavelengths for the $6s ~ {^2}S_{1/2}-6p ~ {^2}P_{1/2,3/2}$ transitions of Cs using circularly and linearly polarized light}

\author{Sukhjit Singh$^a$\footnote{Email: sukhjitphy.rsh@gndu.ac.in}, Kiranpreet Kaur$^a$, B. K. Sahoo$^{b}$
and Bindiya Arora$^a$\footnote{Email: bindiya.phy@gndu.ac.in} }
\affiliation{$^a$Department of Physics, Guru Nanak Dev University, Amritsar, Punjab-143005, India}
\affiliation{$^b$Theoretical Physics Division, Physical Research Laboratory, Navrangpura, Ahmedabad-380009, India}
\date{Received date; Accepted date}
 
\begin{abstract}
We demonstrate magic wavelengths, at which external electric field produces null differential Stark shifts, for the $6s ~ {^2}S_{1/2}-6p ~ 
{^2}P_{1/2,3/2}$ transitions in the Cs atom due to circularly polarized light. In addition, we also obtain magic wavelengths using linearly polarized light, in order to 
verify the previously reported values, and make a comparative study with the values obtained for circularly polarized 
light. A number of these wavelengths are found to be in the optical region and could be of immense interest to experimentalists 
for carrying out high precision measurements. To obtain these wavelengths, we have calculated dynamic dipole polarizabilities 
of the ground, $6p ~{^2}P_{1/2}$ and $6p ~{^2}P_{3/2}$ states of Cs. We use the available precise values of the electric dipole (E1) 
matrix elements of the transitions that give the dominant contributions from the lifetime measurements of the excited states. 
Other significantly contributing E1 matrix elements are obtained by employing a relativistic coupled-cluster singles and doubles
method. The accuracies of the dynamic polarizabilities are substantiated by comparing the static polarizability values with the 
corresponding experimental results.
\end{abstract}

\maketitle

\section{Introduction}
Techniques involving laser cooling and trapping of neutral atoms are of immense interest for many scientific applications, including
those that are capable of probing new physics \cite{phillip} and searching for exotic quantum phase transitions using ultracold atoms ~\cite{grim}.
In particular, trapping atoms using optical lattices have many advantages since they offer long storage
time  ~\cite{grim,balykin,sahoo1} and their energy levels can be easily accessed using lasers \cite{adams}. It is conducive to carry out 
measurements in a transition of an optically trapped atom without realizing the Stark shifts due to the applied laser field. One can achieve 
this by trapping the atom at the wavelengths for which the differential Stark shift of the transition gets nullified. These wavelengths are 
popularly known as magic wavelengths (${\rm \lambda_{magic}}$s) ~\cite{katori}. They play crucial role in state-insensitive quantum 
engineering to set-up many high precision experiments. They are useful in the atomic clock experiments to acquire relative uncertainty 
as small as $\sim \rm 10^{-18}$ ~\cite{porto,sac,sahoo1}, in the quantum information and communication studies
\cite{Monroe}, investigating fundamental physics \cite{fortier,Weiss}, so on. 

Alkali atoms are mostly preferred for performing experiments using laser cooling and trapping techniques. The reason being that the low-lying 
transitions in these atoms are conveniently accessible by the available lasers. Since Cs atom has wider hyperfine splittings in its ground state, 
it has been considered for making microwave clock and for quantum computation. For laser cooling of these atoms, its $6s ~{^2}S_{1/2}$-$6p ~{^2}P_{3/2}$ 
transitions are mainly being used. Owing to a large number of applications of trapped Cs atoms, it would be imperative to find out all plausible 
magic wavelengths of these transitions such that lasers can be appropriately chosen at these wavelengths to reduce systematics significantly in the 
above possible measurements due to the Stark shifts. Mckeever \textit{et al.} had experimentally demonstrated 
${\rm \lambda_{magic}}$ at 935.6 nm for the $6s ~{^2}S_{1/2}$-$6p ~{^2}P_{3/2}$ transition in Cs \cite{mck} using linearly polarized light. Following 
this, magic wavelengths for many alkali atoms including Cs atom were determined using linearly polarized light by Arora \textit{et al.}~\cite{arora1} 
by evaluating the dynamic polarizabilities of these atoms using the relativistic coupled cluster (RCC) method. The estimation of ac Stark shifts using 
circularly polarized light may be advantageous to look for magic wavelengths owing to the predominant role played by the vector polarizabilities. Since 
the vector polarizability contribution is absent in the use of linearly polarized light, this can open-up new windows to manipulate the magic 
wavelengths for a wide range of applications. Recently, we had investigated magic wavelengths in the lighter alkali atoms for circularly polarized light and 
realized many possible magic wavelengths, especially for the $ns~{^2}S_{1/2}$-$np~{^2}P_{3/2}$ transitions with the ground state principal quantum
number $n$ \cite{sahoo1,arorab,arora2}. However, magic wavelengths for circularly polarized light in the Cs atom have not been explored sufficiently. 

\begin{table*}
\caption{\label{pol1}Contributions from different E1 matrix elements ($d$) to the static polarizabilities of the $6S_{1/2}$, $6P_{1/2}$ and 
$6P_{3/2}$ states of Cs atom. The final results are compared with the previously estimated and available experimental results. Uncertainties are 
given in the parentheses.}
\begin{center}
\begin{tabular}{|ccc|ccc|cccc|}
\hline 
\hline
\multicolumn{3}{|c|}{$6S_{1/2}$ state}  & \multicolumn{3}{c|}{$6P_{1/2}$ state}  &  \multicolumn{4}{c|}{$6P_{3/2}$ state}\\
& & & & & & & & & \\ 
 Transition         &  $d$  &  $\alpha^{(0)}$ & Transition & $d$ & $\alpha^{(0)}$ &  Transition  &   $d$ &   $\alpha^{(0)}$ & $\alpha^{(2)}$ \\
 \hline
 & & & & & & & & & \\ 
 $6S_{1/2}-6P_{1/2}$  &  4.489(7)     & 131.88(2) & $6P_{1/2}-6S_{1/2}$  &   4.489(7)    & -131.88(2) & $6P_{3/2}-6S_{1/2}$       &   6.324(7)    & -124.69(2)  &   124.69(2)             \\
 $6S_{1/2}-7P_{1/2}$  &  0.30(3)     & 0.30      & $6P_{1/2}-7S_{1/2}$  &   4.236(21)   & 178.43(7)  & $6P_{3/2}-7S_{1/2}$       &   6.47(3)     &  225.3(1)   &  -225.3(1)          \\
 $6S_{1/2}-8P_{1/2}$  &  0.09(1)      & 0.02      & $6P_{1/2}-8S_{1/2}$  &   1.0(1)      & 5.86(1)    & $6P_{3/2}-8S_{1/2}$       &   1.5(1)      &  6.21(2)    &  -6.21(2)      \\
 $6S_{1/2}-9P_{1/2}$  &  0.04      & $\sim$0   & $6P_{1/2}-9S_{1/2}$  &   0.55(6)     & 1.41       & $6P_{3/2}-9S_{1/2}$       &    0.77(8)     &  1.44   &  -1.44        \\ 
 $6S_{1/2}-10P_{1/2}$ &  0.02      & $\sim$0   & $6P_{1/2}-10S_{1/2}$ &   0.36(4)     & 0.57       & $6P_{3/2}-10S_{1/2}$      &    0.51(5)     &  0.57  &  -0.57     \\ 
 $6S_{1/2}-11P_{1/2}$ &  0.02       & $\sim$0   & $6P_{1/2}-11S_{1/2}$ &   0.27(3)     & 0.29       & $6P_{3/2}-11S_{1/2}$      &    0.37(4)     &  0.29  &  -0.29      \\  
 $6S_{1/2}-12P_{1/2}$ &  0.01         & $\sim$0   & $6P_{1/2}-12S_{1/2}$ &   0.20(2)     & 0.17       & $6P_{3/2}-12S_{1/2}$      &    0.29(3)     &  0.17  &  -0.17       \\       
 $6S_{1/2}-6P_{3/2}$  &  6.324(7)      & 249.38(3) & $6P_{1/2}-5D_{3/2}$  &   7.016(24)   & 1084.3(5)  & $6P_{3/2}-5D_{3/2}$       &    3.166(16)    &  132.51(4)  &   106.01(3)                 \\
 $6S_{1/2}-7P_{3/2}$  &  0.60(6)         & 1.20      & $6P_{1/2}-6D_{3/2}$  &   4.3(4)      & 120.98(9)  & $6P_{3/2}-6D_{3/2}$       &   2.1(2)      &  15.54(7)   &   12.43(5)             \\
 $6S_{1/2}-8P_{3/2}$  &  0.23(2)      & 0.15      & $6P_{1/2}-7D_{3/2}$  &   2.1(2)      & 21.03(8)   & $6P_{3/2}-7D_{3/2}$       &   1.0(1)      &  2.47   &   1.97         \\ 
 $6S_{1/2}-9P_{3/2}$  &  0.13(1)         & 0.05      & $6P_{1/2}-8D_{3/2}$  &   1.3(1)      & 7.43(1)    & $6P_{3/2}-8D_{3/2}$       &   0.61(6)     &  0.84   &   0.67       \\ 
 $6S_{1/2}-10P_{3/2}$ &  0.09(1)      & 0.02      & $6P_{1/2}-9D_{3/2}$  &   0.93(9)     & 3.55       & $6P_{3/2}-9D_{3/2}$       &    0.43(4)     &  0.40  &   0.32     \\
 $6S_{1/2}-11P_{3/2}$ &  0.06(1)         & 0.01      & $6P_{1/2}-10D_{3/2}$ &   0.71(7)     & 2.01       & $6P_{3/2}-10D_{3/2}$      &    0.33(3)     &  0.22  &   0.18       \\
 $6S_{1/2}-12P_{3/2}$ &  0.05         & 0.01      &                      &               &            & $6P_{3/2}-5D_{5/2}$       &    9.59(8)     &  1174(2)    &  -234.9(4)                \\ 
                      &               &           &                      &               &            & $6P_{3/2}-6D_{5/2}$       &   6.3(6)      &  132(2)     &  -26.4(3) \\
                      &               &           &                      &               &            & $6P_{3/2}-7D_{5/2}$       &   2.9(3)      &  21.6(1)    &  -4.32(3) \\                                                                                                 
                      &               &           &                      &               &            & $6P_{3/2}-8D_{5/2}$       &    1.8(2)      &  7.46(3)    &  -1.49  \\                                      
                      &               &           &                      &               &            & $6P_{3/2}-9D_{5/2}$       &   1.3(1)      &  3.53   &  -0.71 \\                                      
                      &               &           &                      &               &            & $6P_{3/2}-10D_{5/2}$      &   1.0(1)      &  1.98   &  -0.40 \\                                                                                                        
  Main($\rm \alpha_v^{v}$)  &               &  383.03(4) &  Main($\rm \alpha_v^{v}$)                    &               & 1294.2(5)  &  Main($\rm \alpha_v^{v}$)                          &                &  1602(3)   &  -255.9(5)               \\
  Tail($\rm \alpha_v^{v}$)  &               & 0.15(8)   &  Tail($\rm \alpha_v^{v}$)                    &               & 24(12)     &  Tail($\rm \alpha_v^{v}$)                         &                &  25(13)     &  -5(2)            \\
  $\rm \alpha_v^{cv}$    &               & -0.47(0)  &  $\rm \alpha_v^{cv}$                    &               & $\sim$0    &  $\rm \alpha_v^{cv}$                         &                & $\sim$0     &   $\sim 0$          \\
$\rm \alpha_0^{c}$    &               & 16.8(8)   &    $\rm \alpha_0^{c}$                  &               & 16.8(8)    &   $\rm \alpha_0^{c}$                         &                & 16.8(8)     &                  \\
 Total   &               & 399.5(8)  &      Total                 &               & 1335(12)   &       Total                     &                & 1644(13)    &  -261(2)                  \\
& & & & & & & & & \\
 Others    &               & 399.9\cite{derevianko} &   Others                   &               & 1338\cite{arora1}        &    Others           
&                & 1650\cite{arora1}      &   -261\cite{arora1} \\
    &               & 399\cite{borschevsky}  &                      &               & 1290\cite{wijngaarden}   &                           &                & 1600\cite{wijngaarden}    &  -233 \cite{wijngaarden}                 \\
& & & & & & & & & \\
    Experiment   &               & 401.0(6)\cite{amini}  &   Experiment                   &               & 1328.4(6)\cite{hunter}   &  Experiment                        &                & 1641(2)\cite{tanner}    &  -262(2)\cite{tanner}                  \\
\hline
\hline
\end{tabular}
\end{center}
\end{table*} 

In this paper, we determine the magic wavelengths for the 6$s~{^2}S_{1/2}$-6$p~{^2}P_{1/2,3/2}$ transitions in the Cs atom using circularly 
polarized light. For this purpose, we calculate the dipole polarizabilities of the ground, $6p~{^2}P_{1/2}$ and $6p~{^2}P_{3/2}$ states 
very precisely. The static polarizability values are compared with the experimental results to verify accuracies in our results. We also 
determine magic wavelengths due to linearly polarized light for the 6$s~{^2}S_{1/2}$-6$p~{^2}P_{1/2,3/2}$ transitions using these polarizabilities 
and compare with the previously reported values in order to validate our approach. We have used atomic unit (a.u.), unless stated 
otherwise.   

\section{Theory and Method of Evaluation}

In the time independent second order perturbation theory, the Stark shift in the energy of $v^{th}$ level of an atom placed in a static electric field
($\mbox{\boldmath$\cal E$}$) is expressed as \cite{bonin}
\begin{eqnarray}
\label{eq1}
\Delta E_v=\sum_{k\neq v}\frac{|\langle\psi _v|V|\psi _k\rangle|^2}{E_v^0-E_k^0},
\end{eqnarray}
where $V=-\mbox{\boldmath$D$} \cdot \mbox{\boldmath$\cal E$}$ is the perturbing electric-dipole interaction Hamiltonian, $E_i^0$ refers to 
the unperturbed energy of the corresponding level denoted by $i=k,v$ and states with subscript $k$ are the intermediate states to which 
transition from the $v^{th}$ state is allowed by the dipole selection rules. For convenience, Eq. (\ref{eq1}) is simplified to get
\begin{eqnarray}
\label{eq3}
\Delta E_v=-\frac{1}{2}\alpha_v{\cal E}^2,
\end{eqnarray}
where, $\alpha_v$ is the static dipole polarizability, and is given by
\begin{eqnarray}
\label{eq4}
\alpha_v=-2\sum_{k\neq v}\frac{(p^*)_{vk}(p)_{kv}}{\delta E_{vk}}.
\end{eqnarray}
Here, $\delta E_{vk}=E_v^0-E_k^0$ and $(p)_{kv}=\langle \psi _k|D|\psi _v\rangle$ is the electric dipole (E1) matrix element between the 
states $|\psi _v\rangle$ and $|\psi _k\rangle$. Since in a number of applications, oscillating electric fields are used, the above expression
is slightly modified for that case, with polarizability as a function of frequency of the electric field, as\cite{manakov}
\begin{eqnarray}
\label{eq5}
\alpha_v(\omega)=-\sum_{k\neq v}{(p^*)_{vk}(p)_{kv}}\left[\frac{1}{\delta E_{vk}+\omega}+\frac{1}{\delta E_{vk}-\omega}\right] .
\end{eqnarray}
For circularly polarized light, in the absence of magnetic field, the above expression is further parameterized in terms of ranks 0, 1 and 2 
components of the tensor products that are known as scalar ($\rm \alpha_v^{(0)}$), vector ($\rm \alpha_v^{(1)}$) and tensor ($\rm \alpha_v^{(2)}$) 
polarizabilities respectively i.e.  
\begin{eqnarray}
\label{eq}
\alpha_v(\omega)=\alpha_v^{(0)}+\frac{Am_j}{2j_v}\alpha_v^{(1)}-\frac{3m_j^2-j_v(j_v+1)}{2j_v(2j_v-1)}\alpha_v^{(2)},
\end{eqnarray}
where $j_v$ is the angular momentum, $m_j$ is its magnetic projection, and  
\begin{eqnarray}
\alpha_v^{(0)}&=&\frac{1}{3(2j_v+1)}\sum_{j_k}|\langle j_v||D||j_k \rangle|^2 \nonumber \\
              & &\times\left[\frac{1}{\delta E_{kv}+\omega}+\frac{1}{\delta E_{kv}-\omega}\right], \\
\alpha_v^{(1)}&=&-\sqrt{\frac{6j_v}{(j_v+1)(2j_v+1)}}\sum_{j_k}(-1)^{j_k+j_v+1}  \nonumber \\
              & & \times \left\{ \begin{array}{ccc}
                             j_v& 1 & j_v\\
                          1 & j_k &1 
                         \end{array}\right\} |\langle j_v||D|| j_k \rangle|^2 \nonumber \\
              & &\times \left[\frac{1}{\delta E_{kv}+\omega}-\frac{1}{\delta E_{kv}-\omega}\right] 
\end{eqnarray}
and
\begin{eqnarray}
\alpha_v^{(2)}&=&-2\sqrt{\frac{5j_v(2j_v-1)}{6(j_v+1)(2j_v+3)(2j_v+1)}} \nonumber \\ 
& & \times \sum_{j_k}(-1)^{j_k+j_v+1}
                                  \left\{ \begin{array}{ccc}
                                            j_v& 2 & j_v\\
                                            1 & j_k &1 
                                           \end{array}\right\} |\langle j_v||D||j_k \rangle|^2 \nonumber\\
                                         & & \times \left[\frac{1}{\delta E_{kv}+\omega}+\frac{1}{\delta E_{kv}-\omega}\right],
\end{eqnarray} 
for the reduced matrix element $|\langle j_v||D|| j_k \rangle|$ with the angular momentum $j_k$ of the intermediate state $k$.
Here, $A$ is the degree of circular polarization assuming the quantization axis to be in the direction of wave vector and possess values $1$ and $-1$ 
for the right handed and left handed circularly polarized light, respectively. The expressions denoted within curly brackets are 
the angular momentum coupling 6-j symbols \cite{edmonds}.  

For linearly polarized light, “the degree of circular polarization $A=0$”.  and total frequency dependent polarizability in this case can be formulated as 
\begin{eqnarray}
\label{eqlinear}
\alpha_v(\omega)=\alpha_v^{(0)}+\frac{3m_j^2-j_v(j_v+1)}{j_v(2j_v-1)}\alpha_v^{(2)},
\end{eqnarray}
where we assume the quantization axis along the direction of polarization vector.

\begin{figure}
\centering 
\includegraphics[width=\columnwidth,keepaspectratio]{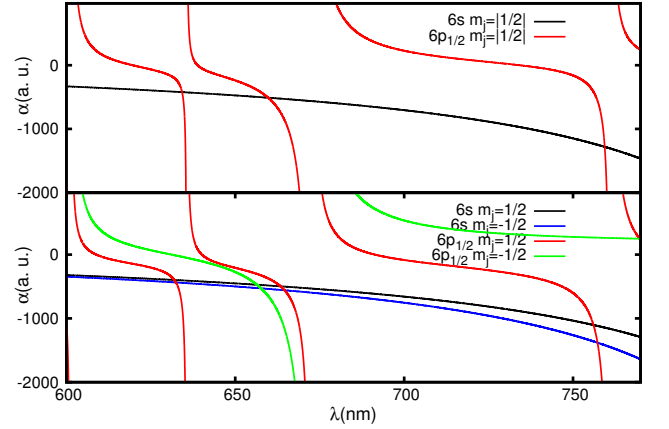}
\caption{(Color online) Dynamic polarizabilities (in a.u.) for the $6S_{1/2}$ and $6P_{1/2}$ states of Cs with different $m_j$
values in the wavelength range 600-770 nm for linearly polarized light (upper half) and circularly polarized light (lower half).}
\label{magiccirc4}
\end{figure}

\begin{figure}
\centering 
\includegraphics[width=\columnwidth,keepaspectratio]{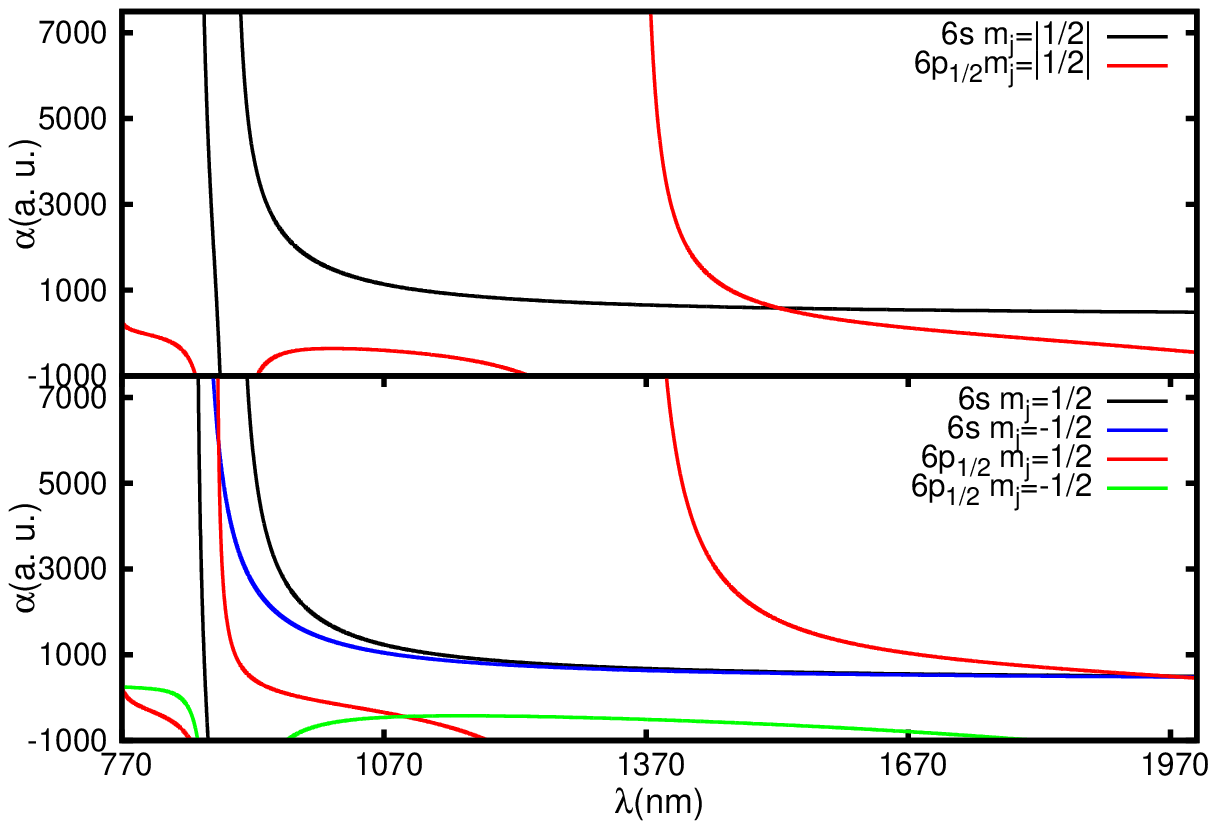}
\caption{(Color online) Dynamic polarizabilities (in a.u.) for the $6S_{1/2}$ and $6P_{1/2}$ states in Cs with different $m_j$
values in the wavelength range 770-2000 nm for linearly polarized light (upper half) and circularly polarized light (lower half).}
\label{magiccirc3}
\end{figure} 

\begin{figure}
\centering 
\includegraphics[width=\columnwidth,keepaspectratio]{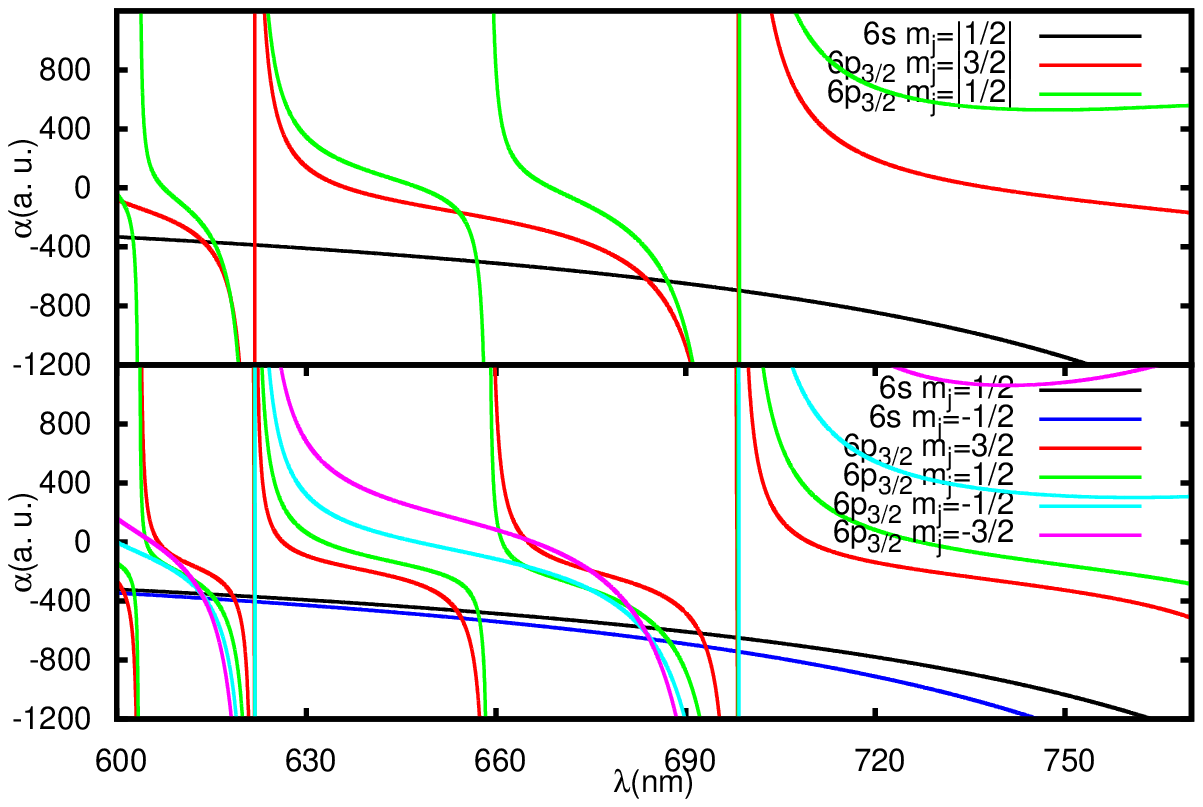}
\caption{(Color online) Dynamic polarizabilities (in a.u.) of the $6S_{1/2}$ and $6P_{3/2}$ states in Cs with different $m_j$
values in the wavelength range 600-770 nm for linearly polarized light (upper half) and circularly polarized light (lower half).}
\label{magiccirc1}
\end{figure}

\begin{figure}
\centering 
\includegraphics[width=\columnwidth,keepaspectratio]{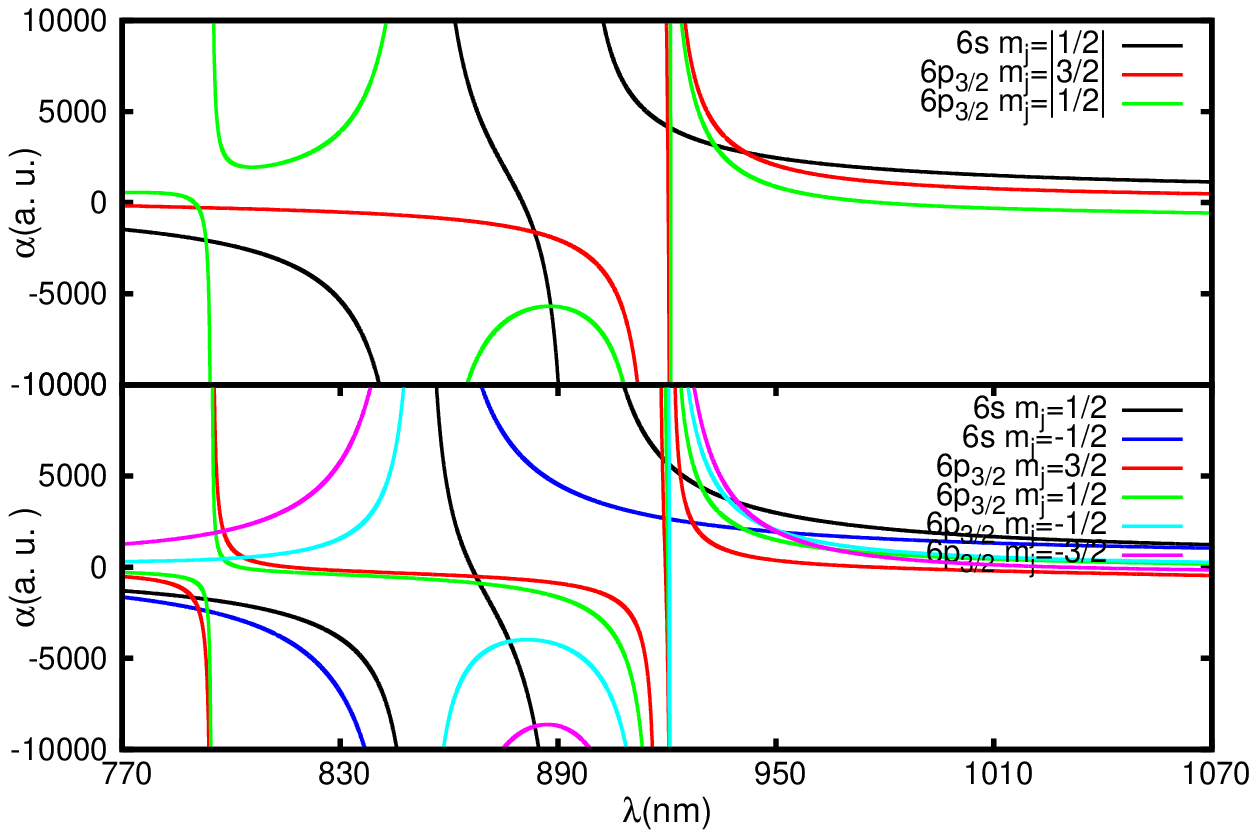}
\caption{(Color online) Dynamic polarizabilities (in a.u.) of the $6S_{1/2}$ and $6P_{3/2}$ states in Cs with different $m_j$ values in the 
wavelength range 770-1070 nm for linearly polarized light (upper half) and circularly polarized light (lower half).}
\label{magiccirc2}
\end{figure}   

\begin{table*}
\caption{\label{magic3}Magic wavelengths ($\lambda _{\rm {magic}}$s) (in nm) with corresponding polarizabilities ($\alpha_v(\omega)$s) (in a.u.)
for the $6S-6P_{1/2}$ transition in the Cs atom with linearly and circularly polarized light along with the resonant wavelengths 
($\lambda_{\rm{res}}$s) (in nm).}
\begin{tabular}{cccccccccccccccccccc}
\hline 
\hline
&&\multicolumn{4}{c}{Linearly Polarization}&\multicolumn{8}{c}{Circularly Polarization}\\ 
 \hline
  & & \multicolumn{2}{c}{Present} & \multicolumn{2}{c}{Ref.\cite{arora1}}  &  \multicolumn{4}{c}{Transition: $6S(m_j=1/2)-6P_{1/2}$}& \multicolumn{4}{c}{Transition: $6S(m_j=-1/2)-6P_{1/2}$}\\
  \cline{3-4} \cline{5-6}\cline{7-14} \\
   & &\multicolumn{4}{c}{$m_j=|1/2|$}&\multicolumn{2}{c}{$m_j=1/2$}&\multicolumn{2}{c}{$m_j=-1/2$}&\multicolumn{2}{c}{$m_j=1/2$}&\multicolumn{2}{c}{$m_j=-1/2$}\\
Resonance &  $\lambda_{\rm{res}}$   &   $\lambda _{\rm {magic}}$   &   $\alpha_v(\omega)$ &   $\lambda _{\rm {magic}}$   &   $\alpha_v(\omega)$ &  $\lambda _{\rm {magic}}$   &   $\alpha_v(\omega)$ & $\lambda _{\rm {magic}}$   &   $\alpha_v(\omega)$  &  $\lambda _{\rm {magic}}$   &   $\alpha_v(\omega)$ & $\lambda _{\rm {magic}}$   &   $\alpha_v(\omega)$ \\
\hline
 & & & & & & & \\
$6P_{1/2}-8D_{3/2}$ &601.22 &&&&&&&\\ 
&& 634.3(2) & -424 & 634.3(2) & -424(2) & 632.0(6)  & -398    &  &  & 632.4(4)  & -437    &  &   \\
$6P_{1/2}-9S_{1/2}$&635.63 &&&&&&&\\ 
&& 659.8(8)    & -511 & 660.1(6) & -513(3) & 663.4(8)  & -497   & 656.4(8) & -473 & 664.6(6)  & -559   & 657.7(7) & -530    \\
$6P_{1/2}-7D_{3/2}$&672.51 &&&&&&&\\
&& 759.38(5)  & -1282 & 759.40(3) & -1282(3)   & 756.1(1) & -1104  & & & 757.25(8) & -1381  & &   \\ 
$6P_{1/2}-8S_{1/2}$&761.10 &&&&&&&\\
$6P_{1/2}-6D_{3/2}$ &876.38 &&&&&&&\\
&&&&&&&&&& 880.79(8) & 5907 &  & \\
$6P_{1/2}-6S_{1/2}$ &894.59 &&&&&&&\\
$6P_{1/2}-7S_{1/2}$ &1359.20 &&&&&&&\\          
&& 1522(3)  &  582 & 1520(3) & 583(2)  &  1966(10)     &  500      &    &  &  1981(10)     &  480      &    &  \\
$6P_{1/2}-5D_{3/2}$ &3011.15 &&&&&&&\\          
\hline
\hline 
\end{tabular}
\end{table*}

The differential ac Stark shift of a transition between the ground state and an excited state is the difference between the ac Stark shifts of the 
two states and is given by
\begin{eqnarray}\nonumber
\delta (\Delta E)_{ge} (\omega ) &=&\Delta E_g(\omega)-\Delta E_e (\omega) \\
                      &=&-\frac{1}{2}\left[\alpha_g(\omega)-\alpha_e(\omega)\right]{\cal E}^2 .
\end{eqnarray}  
Here, subscripts `$g$' and `$e$' represent the ground and excited states, respectively. Our aim is to find out the $\omega$ values at which
$\delta (\Delta E)_{ge} (\omega )$ will be zero. 

\begin{table*}
\caption{\label{magic1} Magic wavelengths ($\lambda _{\rm {magic}}$s) (in nm) with corresponding polarizabilities ($\alpha_v(\omega)$s) (in a.u.) 
for the $6S-6P_{3/2}$ transition in the Cs atom with linearly polarized light along with the resonant wavelengths ($\lambda_{\rm{res}}$s) (in nm).}
\begin{tabular}{cccccccccc}
\hline 
\hline
  \multicolumn{2}{c}{Transition}&\multicolumn{4}{c}{$m_j=|1/2|$}&\multicolumn{4}{c}{$m_j=|3/2|$}\\
   \cline{1-6} \cline{7-10} \\
\multicolumn{2}{c}{$6S(m_j=|1/2|)-6P_{3/2}$}&\multicolumn{2}{c}{Present}&\multicolumn{2}{c}{Ref.~\cite{arora1}}&\multicolumn{2}{c}{Present}&\multicolumn{2}{c}{Ref.~\cite{arora1}}\\
\cline{1-6} \cline{7-10}\\
Resonance &  $\lambda_{\rm{res}}$   &   $\lambda _{\rm {magic}}$   &   $\alpha_v(\omega)$ &  $\lambda _{\rm {magic}}$   &   $\alpha_v(\omega)$  &   $\lambda _{\rm {magic}}$   &   $\alpha_v(\omega)$ &  $\lambda _{\rm {magic}}$   &   $\alpha_v(\omega)$ \\
\hline
$6P_{3/2}-9D_{5/2}$ &584.68 &&&\\ 
&& 602.6(3)  & -338     &  602.6(4) & -339(1)      &   &  &  &      \\ 
$6P_{3/2}-10S_{1/2}$&603.58 &&&\\ 
&& 615.4(10)  & -370  & 615.5(8) & -371(3)   & 614(1) & -365   & 614(3) & -367(8)\\
$6P_{3/2}-8D_{5/2}$&621.48 &&&\\
&& 621.924(4)  & -387 & 621.924(2) & -388(1)    & 621.85(4) & -387 & 621.844(3) & -388(1) \\  
$6P_{3/2}-8D_{3/2}$&621.93 &&&\\
&& 657.0(2)  & -500  & 657.05(9) & -500(1)  &       &   &     &\\
$6P_{3/2}-9S_{1/2}$ &658.83 &&&\\
&& 687(1)  & -633 & 687.3(3) & -635(3)   & 684(1) & -617 & 684.1(5) & -618(4)  \\  
$6P_{3/2}-7D_{5/2}$ &697.52 &&&\\
&& 698.5(5)  & -697   & 698.524(2) & -697(2) & 698.3(7) & -695 & 698.346(4) & -696(2)     \\
$6P_{3/2}-7D_{3/2}$ &698.54 &&&\\          
&& 793.1(2)  & -2072  & 793.07(2) & -2074(5) &       &     &  &   \\  
$6P_{3/2}-8S_{1/2}$ &794.61 &&&\\
$6S_{1/2}-6P_{3/2}$ &852.35 &&&\\
&&  888(2) & -5690  & 887.95(10) & -5600(100) &   884(3) & -1618 & 883.4(2) & -1550(90)     \\
$6S_{1/2}-6P_{1/2}$ &894.59 &&&\\
$6P_{3/2}-6D_{5/2}$ &917.48 &&&\\       
&&  921.0(9)    &  4088   & 921.01(3)       &     4088(10)    &  920(2)    & 4180 & 920.18(6) &  4180(14) \\   
$6P_{3/2}-6D_{3/2}$ &921.11 &&&\\ 
&&  933(8)  &  3153 & 932.4(8) &  3197(50) & 941.7(3) &  2752 & 940.2(1.7) &  2810(70)  \\        
\hline
\hline 
\end{tabular}
\end{table*}

\begin{table*}
\caption{\label{magic2} Magic wavelengths ($\lambda _{\rm {magic}}$s) (in nm) with corresponding polarizabilities ($\alpha_v(\omega)$s) (in a.u.) for 
the $6S(m_j=1/2)-6P_{3/2}$ transition in the Cs atom with the circularly polarized light (A=-1) along with the resonant wavelengths 
($\lambda_{\rm{res}}$s) (in nm).}
\begin{center}
\begin{tabular}{ccccccccccccccccc}
\hline 
\hline
 \multicolumn{2}{c}{Transition: $6S(m_j=1/2)-6P_{3/2}$}&\multicolumn{2}{c}{$m_j=3/2$}&\multicolumn{2}{c}{$m_j=1/2$}&\multicolumn{2}{c}{$m_j=-1/2$}&\multicolumn{2}{c}{$m_j=-3/2$}\\
 \hline
Resonance &  $\lambda_{\rm{res}}$   &   $\lambda _{\rm {magic}}$   &   $\alpha_v(\omega)$ &  $\lambda _{\rm {magic}}$   &   $\alpha_v(\omega)$ & $\lambda _{\rm {magic}}$   &   $\alpha_v(\omega)$ &  $\lambda _{\rm {magic}}$   &   $\alpha_v(\omega)$  \\
\hline
$6P_{3/2}-9D_{5/2}$ &584.68 &&&&&&&&\\ 
&& 600.9(9) & -323  & 602.7(7)  & -327    &  &   &   &   \\ 
$6P_{3/2}-10S_{1/2}$&603.58 &&&&&&&&\\ 
&& 618(1)    & -363  & 615(2)  & -354    & 613(2) & -351  & 613(1) &  -351  \\
$6P_{3/2}-8D_{5/2}$&621.48 &&&&&&&&\\
&&  622(1) & -371      & 621.8(5) & -371 & 621.9(5)    & -371    &   &  \\  
$6P_{3/2}-8D_{3/2}$&621.93 &&&&&&&&\\
&& 654(1)  & -464    & 657.2(3)      & -475    &  &     &  & \\
$6P_{3/2}-9S_{1/2}$ &658.83 &&&&&&&&\\
&& 692(2)  & -619    & 686(2) & -588  &  683(2)  & -576  & 683(1) & -577  \\ 
$6P_{3/2}-7D_{5/2}$ &697.52 &&&&&&&&\\
&&698.2(8)  & -648  &  698.3(2)  & -648    & 698.4(2)     & -650      &  &   \\
$6P_{3/2}-7D_{3/2}$ &698.54 &&&&&&&&\\          
&& 789.3(5)  & -1661   &  792.9(3)     &  -1749      &    &     &  &  \\  
$6P_{3/2}-8S_{1/2}$ &794.61 &&&&&&&&\\
$6S_{1/2}-6P_{3/2}$ &852.35 &&&&&&&&\\
&&  867(2) & -510 & 868(2) & -891   & 877(2) & -4073    & 884(1) & -8725  \\
$6S_{1/2}-6P_{1/2}$ &894.59 &&&&&&&&\\
$6P_{3/2}-6D_{5/2}$ &917.48 &&&&&&&&\\       
&&  919(2)    &  6000      &  920(2)    & 5761  & 920.6(3)      & 5605    &   & \\   
$6P_{3/2}-6D_{3/2}$ &921.11 &&&&&&&&\\ 
&&  924(3)  &  5116  & 930(6) &  4375 & 937(5)  & 3739   & 940(4) &  3543  \\
\hline
\hline 
\end{tabular}
\end{center}
\end{table*}
  
For a state of an atomic system having a closed core and a valence electron, dipole polarizability can be conveniently evaluated by calculating
contributions separately due to the core, core-valence and valence correlations \cite{nandy, jasmeet, sukhjit}. In other words, we can write
\begin{eqnarray}
\alpha_v(\omega)=\alpha_0^{c}(\omega)+\alpha_v^{cv}(\omega)+\alpha_v^{v}(\omega), 
\end{eqnarray}
where $\alpha_0^{c}(\omega)$, $\alpha_v^{cv}(\omega)$ and $\alpha_v^{v}(\omega)$ are the contributions from the core, core-valence and valence 
correlation effects, respectively. The subscript `$0$' in $\alpha_0^{c}(\omega)$ means that it is independent of the valence orbital in a state.
For estimating the dominant $\alpha_v^{v}(\omega)$ contributions, we calculate the wave functions of many low-lying excited states 
($|\Psi_k\rangle s$) using a linearized version of the RCC method in the singles and doubles approximation (SD method) \cite{blundell,safronova12,
safronova22}. In this method, the wave functions of the ground, $6p~{^2}P_{1/2}$ and $6p~{^2}P_{3/2}$ states in Cs, that have a common core 
$[5p^6]$, are expressed as
\begin{eqnarray}{\label{wav}}
|\Psi_v\rangle &=& [1+T_1+T_2 +S_{1v}+S_{2v}] |\Phi_v\rangle \nonumber \\
        &=&[1+\sum_{ma}\rho_{ma}a_m^\dagger a_a +  \frac{1}{2}\sum_{nmab}\rho_{mnab}a_m^\dagger a_n^\dagger a_ba_a \nonumber \\
& &+\sum_{m\neq v}\rho_{mv}a_m^\dagger a_v+\sum_{mna}\rho_{mnva}a_m^\dagger a_n^\dagger a_aa_v]|\Phi_v\rangle, \ \ \ \ \
\end{eqnarray}
where $T_1$ and $T_2$ are the singles and doubles excitation operators, respectively, that are responsible for exciting only the core electrons
from $|\Phi_v\rangle$, while $S_{1v}$ and $S_{2v}$ are the singles and doubles excitation operators, respectively, that excite valence electron along 
with other core electrons from $|\Phi_v\rangle$ as described by the second quantized creation operator $a^\dagger$ and annihilation operator $a$ with 
the appropriate subscripts. Indices $m$, $n$ and $r$ refer to the virtual electrons, 
indices $a$ and $b$ represent the core electrons and $v$ corresponds to the valence electron. The coefficients $\rho_{ma}$ and $\rho_{mv}$ are the 
singles and doubles excitation amplitudes involving the core electrons alone while $\rho_{mnab}$ and $\rho_{mnva}$ refer to the singles and 
doubles excitation amplitudes involving the valence orbital $v$ from $|\Phi_v\rangle$. We obtain $|\Phi_v\rangle$ by expressing
\begin{equation}
|\Phi_v\rangle=a_v^\dagger|\Phi_0\rangle,
\end{equation}
where $|\Phi_0\rangle$ is the Dirac-Hartree-Fock (DHF) wave function of a closed core $[5p^6]$.

 The E1 matrix element for a transition between the states $|\Psi_v \rangle$ and $|\Psi_k\rangle$ are calculated using the expression
\begin{eqnarray}
D_{vk}&=&\frac{\langle\Psi_v|D|\Psi_k\rangle}{\sqrt{\langle\Psi_v|\Psi_v\rangle\langle\Psi_k|\Psi_k\rangle}} \nonumber \\
      &=& \frac{\langle\Phi_v|\tilde{D}|\Phi_k\rangle}{\sqrt{\langle\Phi_v|\{1+\tilde{N}_v\}|\Phi_v\rangle\langle\Phi_k|\{1+\tilde{N}_k\}|\Phi_k\rangle}} ,
\end{eqnarray}
where $\tilde{D}=\{1+S_{1v}^{\dagger}+S_{2v}^{\dagger}+T_1^{\dagger}+T_2^{\dagger}\}D\{1+S_{1k}+S_{2k}+T_1+T_2\}$ and 
$\tilde{N}_v=\{S_{1v}^{\dagger}+S_{2v}^{\dagger}+T_1^{\dagger}+T_2^{\dagger}\}\{S_{1v}+S_{2v}+T_1+T_2\}$. For practical purposes, we calculate 
the E1 matrix elements of low-lying transitions, which contribute dominantly to $\alpha_v^v$, and refer to the result as ``Main($\alpha_v^{v}$)''
contribution. Contributions from the other high-lying excited states, including the continuum, are estimated using the DHF method and given as
``Tail($\alpha_v^{v}$)''. We, again, estimate $\alpha_v^{cv}$ and $\alpha_0^c$ contributions using the DHF method.

\section{Results and Discussion}

To find precise values of ${\rm \lambda_{magic}}$s for the $6S-6P_{1/2,3/2}$ transitions in the Cs atom, accurate values of the dynamic dipole 
polarizabilities of the involved states are prerequisites. To evince the accuracies of these results, we first evaluate the static
polarizabilities ($\alpha_v(0)$) of these states and compare them with their respective experimental values and 
previously reported precise calculations. We give both scalar and tensor polarizabilities of the considered ground, $6p ~{^2}P_{1/2}$ and 
$6p ~{^2}P_{3/2}$ states of Cs in Table \ref{pol1} using our calculations along with other results. Contributions from ``Main'' and ``Tail'' to 
$\alpha_v^v$, core-valence and core contributions to our calculations are given explicitly in this table. We also tabulate the E1 matrix elements
used for determining the ``Main'' contributions to $\alpha_v^v$. 

\begin{table*}
\caption{\label{magic22} Magic wavelengths ($\lambda _{\rm {magic}}$s) (in nm) with corresponding polarizabilities ($\alpha_v(\omega)$s) (in a.u.) for 
the $6S(m_j=-1/2)-6P_{3/2}$ transition in the Cs atom with the circularly polarized light (A=-1) along with the resonant wavelengths 
($\lambda_{\rm{res}}$s) (in nm).}
\begin{center}
\begin{tabular}{ccccccccccccccccc}
\hline 
\hline
  \multicolumn{2}{c}{Transition: $6S(m_j=-1/2)-6P_{3/2}$}&\multicolumn{2}{c}{$m_j=3/2$}&\multicolumn{2}{c}{$m_j=1/2$}&\multicolumn{2}{c}{$m_j=-1/2$}&\multicolumn{2}{c}{$m_j=-3/2$}\\
 \hline
Resonance &  $\lambda_{\rm{res}}$   &   $\lambda _{\rm {magic}}$   &   $\alpha_v(\omega)$ &  $\lambda _{\rm {magic}}$   &   $\alpha_v(\omega)$ & $\lambda _{\rm {magic}}$   &   $\alpha_v(\omega)$ &  $\lambda _{\rm {magic}}$   &   $\alpha_v(\omega)$  \\
\hline
$6P_{3/2}-9D_{5/2}$ &584.68 &&&&&&&&\\ 
&& 601(1) & -349  & 602.8(5)  & -353    &  &   &   &   \\ 
$6P_{3/2}-10S_{1/2}$&603.58 &&&&&&&&\\ 
&& 618.7(9)    & -395  & 616(2)  & -386    & 614(1) & -382  & 613.8(9) &  -381  \\
$6P_{3/2}-8D_{5/2}$&621.48 &&&&&&&&\\
&&  621.8(8) & -404      & 621.8(1) & -404 & 621.9(5)    & -404    &   &  \\  
$6P_{3/2}-8D_{3/2}$&621.93 &&&&&&&&\\
&& 654.5(7)  & -517    & 657.4(2)      & -528    &  &     &  & \\
$6P_{3/2}-9S_{1/2}$ &658.83 &&&&&&&&\\
&& 693(2)  & -710    & 687(2) & -675  &  685(1)  & -659  & 684.3(9)  & -657 \\ 
$6P_{3/2}-7D_{5/2}$ &697.52 &&&&&&&&\\
&&698.2(8)  & -742  &  698.3(2)  & -743    & 698.4(2)     & -744      &  &   \\
$6P_{3/2}-7D_{3/2}$ &698.54 &&&&&&&&\\          
&& 791.0(3)  & -2299   &  793.4(9)     &  -2409      &    &     &  &  \\  
$6P_{3/2}-8S_{1/2}$ &794.61 &&&&&&&&\\
$6S_{1/2}-6P_{3/2}$ &852.35 &&&&&&&&\\
&&&&&&&&&   \\
$6S_{1/2}-6P_{1/2}$ &894.59 &&&&&&&&\\
$6P_{3/2}-6D_{5/2}$ &917.48 &&&&&&&&\\       
&&  919(2)    &  2691      &  920.1(8)    & 2658  & 920.7(2)      & 2638    &   & \\   
$6P_{3/2}-6D_{3/2}$ &921.11 &&&&&&&&\\ 
&&  927(4)  &  2437  & 941.0(5) &  2102 & 952(3)  & 1912   & 950(5) &  1932  \\
\hline
\hline 
\end{tabular}
\end{center}
\end{table*}   

To reduce the uncertainties in the evaluation of these polarizability values, we use E1 matrix elements for $6S-6P$ transitions extracted 
from the very precisely measured lifetimes of the $6p ~{^2}P_{3/2}$ and $6p ~{^2}P_{1/2}$ states of Cs by Rafac {\it et al}. \cite{rafac}. We 
also use E1 matrix elements for the $6P-7S$ transitions compiled in Ref. \cite{safronova22}, which are derived from the measured lifetime of 
the $7S$ state. Similarly, the E1 matrix element of $6P_{1/2}-5D_{3/2}$ transition has been derived by combining the measured differential 
Stark shift of the D1 line with the experimental value of the ground state dipole polarizability of Amini {\it et al}. \cite{amini}. We
adopt a procedure similar to that is given in Ref. \cite{Arora3} to determine the E1 matrix elements from the measured Stark shifts. The values 
of these matrix elements along with their experimental uncertainties are listed in Table~\ref{pol1}. Otherwise, the required E1 matrix elements 
for considered transitions up to $12S$, $12P$ and $10D$ states are obtained by employing SD method as described in the previous section. The 
uncertainties in these matrix elements are calculated by comparing matrix elements for the $6S-6P$ and $6P-7S$ transitions calculated using our 
method and available experimental values. The maximum difference between the experiment and our results for these matrix elements is 6\%. 
Therefore, we assign maximum a 10\% uncertainty to all the matrix elements given in Table~\ref{pol1}. We have used 70 B-spline functions confined 
within a cavity of radius $R=220$ a.u. to construct the single-particle orbitals. We use experimental values of the excitation energies of these 
transitions from the National Institute of Science and Technology (NIST) database \cite{nist} to reduce further the uncertainties in the 
evaluation of the polarizabilities.

Our calculated value of $\alpha_v(0)$ for the ground state is 399.5 a.u., which matches very well with other theoretical values 399 a.u. and 399.9 
a.u. estimated by Borschevsky {\it et al.} \cite{borschevsky} and Derevianko {\it et al.} \cite{derevianko}, using the other variants of RCC method, 
respectively. These results are also in very good agreement with the experimental result 401.0(6) a.u. measured by Amini {\it et al.} \cite{amini} 
using the time-of-flight technique. Similarly our calculation gives $\alpha_v(0)$ of the $6P_{1/2}$ state to be 1335 a.u., which is slightly larger 
than the other calculated value 1290 a.u. of Wijngaarden {\it et al.} \cite{wijngaarden} but agrees quite well with another calculated value 1338 a.u. reported by Arora {\it et al.} \cite{arora1} and the measured value 1328.4(6) a.u. reported in Ref. \cite{hunter}. In the work of Wijngaarden {\it et al.}, polarizabilities were evaluated using the oscillator strengths from the method of Bates and Damgaard \cite{bates}. The scalar and tensor 
polarizabilities of the $6P_{3/2}$ state using our method are obtained to be 1644 a.u. and $-261$ a.u., respectively. They are also in very good agreement
with the experimental values reported in Ref. \cite{tanner} and are in reasonable agreement with the theoretical values reported by Arora {\it et al.}
\cite{arora1} and Wijngaarden {\it et al.} \cite{wijngaarden}. The above analysis shows that we have obtained very accurate values of the dipole 
polarizabilities using our method of evaluation. This justifies that determining dynamic polarizabilities using the same procedure can also provide 
competent results. Hence, $\rm \lambda_{magic}$ values of the $6S-6P_{1/2,3/2}$ transitions in Cs can be determined  without any ambiguity
using these accurate values of the dipole polarizabilities. 

We now proceed to determine $\rm \lambda_{magic}$s for the $6S-6P_{1/2,3/2}$ transitions in the Cs atom. For this purpose, we plot 
the dynamic dipole polarizabilities of the $6S$ and $6P_{1/2,3/2}$ states in Figs. \ref{magiccirc4}, \ref{magiccirc3}, \ref{magiccirc1} and 
\ref{magiccirc2}. They are shown in the upper and lower halves for linearly and circularly polarized light respectively. We use left 
circularly polarized light ($A=-1$) while determining the magic wavelengths. Note that it is not required to produce results for right circularly 
polarized light ($A=1$) separately, because the results for $\rm \lambda_{magic}$s will be same as that for left circularly polarized light with  
the counter sign of $m_j$ sublevels. For this purpose we consider only left handed circularly polarized light with all positive and negative 
$m_j$ values. We consider all possible $m_j$ values of the $6S$ and $6P_{1/2,3/2}$ states. It is evident from the above plots 
that for the considered wavelength range, the dynamic polarizability for the 
$6S$ state is generally small except in the close vicinity of the resonant $6S_{1/2}-6P_{1/2}$ and $6S_{1/2}-6P_{3/2}$ transitions. On the other 
hand, polarizabilities of the $6P$ states have significant contributions from several resonant transitions. Thus, the polarizability curves of 
the $6P$ states cross with the polarizability curve of the $6S$ state in between these resonant transitions. The wavelengths at which intersections of these 
polarizability curves take place are recognized as $\rm \lambda_{\rm{magic}}$s in the above figures for both linearly and circularly polarized light. We 
have also tabulated these values for the $6S-6P_{1/2,3/2}$ transitions in Tables \ref{magic3}, ~\ref{magic1}, \ref{magic2} and \ref{magic22} 
along with their respective uncertainties in the parentheses. These uncertainties are estimated by considering maximum possible errors in the 
estimated differential polarizabilities between the involved states in a transition. The corresponding values of the dynamic polarizabilities are 
also mentioned in the above tables to provide an estimate of the kind of trapping potential required at those magic 
wavelengths. We also list the resonant wavelengths ($\lambda_{\rm{res}}$) in the tables to highlight the respective placements of these 
$\lambda_{\rm{magic}}$s.

As seen from Table \ref{magic3}, we find $ \lambda_{\rm{magic}}$s for the $6S(m_j=1/2)-6P_{1/2}$ transition around 630 nm, 660 nm and 760 nm for 
both linearly and circularly polarized light. Other $\lambda_{\rm{magic}}$s are located at 1522 nm for the linearly polarized light and around 
1966 nm with $6P_{1/2}$($m_j$=1/2) for the circularly polarized light. $\lambda_{\rm{magic}}$s at 632, 756.1 and 1966 nms does not support 
state-insensitive trapping for $m_j=-1/2$ sublevel of the $6P_{1/2}$ state and hence switching trapping scheme as discussed in Ref.~\cite{arora2} 
can be used here. In this approach, the change of sign of 
$A$ will lead to the same result for the positive values of $m_j$ sublevels of the $6P$ state. $\lambda_{\rm{magic}}$s at 1522 nm and 1966.1 nm  support the red detuned trap, while the other above mentioned 
$\lambda_{\rm{magic}}$s support the blue detuned traps. Values of $ \lambda_{\rm{magic}}$s for the linearly polarized light are also compared with 
$\lambda_{\rm{magic}}$s of Arora {\it et al.} reported in Ref. \cite{arora1}. Both findings agree with each other, as the method of calculation 
in both is almost similar. Similarly, we tabulate $\lambda_{\rm{magic}}$s for the $6S(m_j=-1/2)-6P_{1/2}$ transition in same table. It 
can be evidently seen from the table that $\lambda_{\rm{magic}}$s are red shifted from the $\lambda_{\rm{magic}}$s for $6S(m_j=1/2)-6P_{1/2}$. We 
have also determined an extra $\lambda_{\rm{magic}}$ at 880.79 nm, which supports a red detuned trap. 

We list $\lambda_{\rm{magic}}$s for the $6S(m_j=1/2)-6P_{3/2}$ transition for linearly and circularly polarized light separately in Tables 
\ref{magic1} and \ref{magic2} respectively. In case of the linearly polarized light, at least ten $\lambda_{\rm{magic}}$s are systematically 
located between the resonant transitions with $m_j=|1/2|$, while only seven $\lambda_{\rm{magic}}$s are located for the $m_j=|3/2|$ sublevel. It, 
thus, implies that use of linearly polarized light does not completely support state insensitive trapping of this transition and results are 
dependent on the $m_j$ sublevels of the $6P_{3/2}$ state. This is also in agreement with the results presented in Ref. ~\cite{arora1}. The 
experimental magic wavelength at 935.6 nm, as demonstrated by McKeever \textit{et. al.} \cite{mck}, matches well with the average of the last two 
magic wavelengths obtained at 933 nm (for $m_j=|1/2|$) and 941.7 nm (for $m_j=|3/2|$). As shown in Table \ref{magic2}, we get a set of ten magic 
wavelengths for circularly polarized light in between the eleven $6P_{3/2}$ resonances lying in the wavelength range 600-1600 nm. Those magic 
wavelengths for which the values corresponding to $m_j=-1/2,-3/2$ sublevels are absent, do not support state-insensitive trapping, and we 
recommend the use of a switching trapping scheme for this transition as proposed in \cite{arora2}.  For the $6S(m_j=-1/2)-6P_{3/2}$ transition, 
we enlist the $\lambda_{\rm{magic}}$s in Table \ref{magic22}. These $\lambda_{\rm{magic}}$s are slightly red shifted to those demonstrated for 
$6S(m_j=-1/2)-6P_{3/2}$ transition. 

\section{Conclusion}
We have investigated possible magic wavelengths within the wavelength range 600 - 2000 nm for the $6S-6P_{1/2,3/2}$ transitions in the Cs atom 
considering both linearly and circularly polarized light. Our values for linearly polarized light were compared with the previously estimated 
values and they are found to be in good agreement. With circularly polarized light, we find a large number of magic wavelengths that are in 
the optical region and would be of immense interest for carrying out many precision measurements at these wavelengths where the above transitions 
are used for the laser cooling purposes. We have used very precise electric dipole matrix elements, extracting from the observed lifetimes 
and evaluating using the relativistic coupled-cluster method, to evaluate the dynamic polarizabilities of the Cs atom very 
precisely. These quantities are used to determine the above magic wavelengths. By comparing static values of the polarizabilities with their 
respective experimental results, accuracies of the polarizabilities and magic wavelengths were adjudged. In few situations, we found it would be 
advantageous to use the magic wavelengths of circularly polarized light over linearly polarized light. As an example, magic wavelengths 
for circularly light are missing for some $m_j=1/2,3/2$ sublevel but they are present for the corresponding $-m_j$ sublevel or vice-versa. In
this case, one can switch the polarization of the light and can successfully locate the positions of the magic wavelengths.   

\section*{Acknowledgements}
The work of S.S. and B.A. is supported by CSIR grant no. 03(1268)/13/EMR-II, India. K.K. acknowledges the financial support from DST (letter 
no. DST/INSPIRE Fellowship/2013/758). The employed SD method was developed in the group of Professor M. S. Safronova of the University of
Delaware, USA.


\end{document}